\documentclass[preprint,showpacs,preprintnumbers,amsmath,amssymb]{revtex4}

\usepackage{graphicx}
\usepackage{dcolumn}
\usepackage{bm}

\begin{document}

\preprint{KEK-TH-1196}

\title{Warped String Compactification via\\
Singular Calabi-Yau Conformal Field Theory}

\author{Shun'ya Mizoguchi}
 \altaffiliation[Also at ]{Department of Particle and Nuclear Physics, The Graduate University for Advanced Studies.}
\author{}%
\affiliation{%
High Energy Accelerator Research Organization (KEK)\\
Tsukuba, Ibaraki 305-0801, Japan 
}%

\author{\phantom{}}
\affiliation{
}%

\date{August 25, 2008\\ \phantom{} }

\begin{abstract}
We construct spacetime supersymmetric, modular invariant partition
functions of strings on the conifold-type singularities which
include contributions from the discrete-series representations of SL(2, R). 
The discrete spectrum is automatically consistent with the GSO 
projection in the continuous sector, and contains massless 
matter fields localized on a four-dimensional submanifold at the tip of a cigar. 
In particular, they are in the $27\oplus 1$ of $E_6$ for the $E_8\times E_8$ heterotic string. 
We speculate about a possible realization of local $E_6$ GUT by using this 
framework.  
\end{abstract}

\maketitle

\section{\label{sec:Introduction}Introduction
}
%

Recent superstring theory is confronted with the problem of the landscape
\cite{Suskind}. 
The problem is twofold. First, one needs to 
stabilize various moduli of a compact Calabi-Yau manifold.
The second part of the problem is that there are many ways to do that, 
and the number of ways is even astronomically large.

In this Letter, we consider instead superstrings on a 
Calabi-Yau three-fold with an isolated singularity.
There are a number of reasons why such singular Calabi-Yau manifolds are 
of interest. First, since only the collapsing cycles are focused on, the number 
of moduli can be small, and the types of singularities are classified in a 
simple way. %
The second reason for the interest is that the noncompact Gepner model 
construction offers a new framework for warped compactification of superstrings.
Finally, the third reason 
is that the set-up, by construction, escapes the no-go theorem against an 
accelerating universe \cite{Gibbons}.

We first construct supersymmetric, modular invariant partition 
functions on the ADE generalizations of the conifold including 
contributions from the discrete series of $SL(2,{\bf R})$. 
Using the character decomposition, we then show that 
there are massless matter fields in the discrete spectrum,
which are hyper/vector multiplets in type IIA/IIB strings, and in the 
$27\oplus 1$ representation of $E_6$ in the $E_8\times E_8$ heterotic case.
This provides a picture of spacetime consisting 
of a warped product of a Minkowski space and a 
two-dimensional Euclidean black hole \cite{Witten}, 
where the massless matter is  
localized \cite{DVV} 
on a four-dimensional submanifold at the cigar tip. 

The chiral ring structure of the $SL(2,{\bf R})/U(1)$ 
Kazama-Suzuki model was already investigated in \cite{ESconifoldtype}.
One of the virtues of the new partition function (eq.(\ref{Z_ADE_new(tau)})) is 
that it enables us to easily determine the Lorentz quantum numbers of these discrete states,
not only for type II but for heterotic strings, which are automatically consistent with the GSO projection in the continuous sector.

To avoid confusion it should be noted that, although our system may 
appear similar to the familiar warped deformed conifold geometry \cite{KSKT},  
there are the following differences: (i) We do not place any D-branes 
in the background conifold geometry. The localized modes are closed string 
modes coming from the geometric moduli of the Calabi-Yau, 
and not the open string modes on D-branes.
(ii) Our picture of warped spacetime arises as an effective geometry of 
the gauged WZW model.  While the whole system is a direct product of 
two (4D spacetime and 2D black hole) CFTs, the effective 6D metric is 
warped in the Einstein frame because of the nontrivial dilaton profile
of the black hole.

This article is a highly compressed version of the report 
on our results.
A detailed account of the material presented here will be 
given in a separate publication \cite{Mizoguchi:Localized}.

\section{Modular invariant partition functions with discrete-series representations}

A noncompact Calabi-Yau threefold with an isolated singularity of the ADE type 
is described \cite{OV} by a tensor product of the $SL(2,{\bf R})/U(1)$ Kazama-Suzuki 
model \cite{KS} 
at level $\kappa=k+2$ and an $N=2$ minimal model at level $k_{\mbox{\scriptsize min}}$.
The central charge of the Kazama-Suzuki model is $c_{\mbox{\scriptsize KS}}=\frac{3(k+2)}k$ and 
that of the minimal model is $
c_{\mbox{\scriptsize min}}=\frac{3 k_{\mbox{\scriptsize min}}}{k_{\mbox{\scriptsize min}}+2}
$. 
They must add up to nine, and hence 
$
k=\frac{2(k_{\mbox{\scriptsize min}}+2)}{k_{\mbox{\scriptsize min}}+4}
$.
Modular invariant partition functions for the Kazama-Suzuki model
with contributions from both the continuous and discrete series of $SL(2,{\bf R})$
were derived in \cite{ES2} by the path-integral approach 
:   
\begin{eqnarray}
Z^{(NS)}(\tau)&=&
\int_0^1 ds_1 \int_0^1 ds_2
\frac{|\vartheta_3(\tau,s_1\tau-s_2)|^2}{|\vartheta_1(\tau,s_1\tau-s_2)|^2}
\sum_{v,w\in {\bf Z}}e^{-\frac{k\pi}{\tau_2}|(w+s_1)\tau-(v+s_2)|^2},
\label{Z_superSL2RoverU1}
\end{eqnarray}
where the expressions for other spin structures can be obtained by an obvious 
replacement of the theta function. By a Poisson resummation we may write
\begin{eqnarray}
\sum_{v,w\in {\bf Z}}e^{-\frac{k\pi}{\tau_2}|(w+s_1)\tau-(v+s_2)|^2}
&=&
\sum_{n,w\in {\bf Z}}
e^{-k \pi \tau_2 s_1^2}
q^{\frac{m^2}k}e^{-2\pi i m (s_1\tau -s_2)}
\bar{q}^{\frac{{\tilde m}^2}k}e^{+2\pi i {\tilde m} (s_1\bar\tau -s_2)},
\label{lattice_decomposition}
\end{eqnarray}
where $m=\frac{n-kw}2$, ${\tilde m}=-\frac{n+kw}2$. They run over an appropriate 
direct sum of orthogonal lattices determined by $n,w\in {\bf Z}$.

Let us first consider the case $k=1$ ($k_{\mbox{\scriptsize min}}=0$) 
which corresponds to the conifold. 
The summation (\ref{lattice_decomposition}) already looks like a product of theta functions
and can be written as 
\begin{equation}
(\ref{lattice_decomposition}) 
=\sum_{\nu\in {\bf Z_2}}
e^{- \pi \tau_2 s_1^2}
\Theta_{\nu,1}(\tau,s_2-s_1\tau)
\overline{\Theta_{\nu,1}(\tau,s_2-s_1\tau)}
\label{thetatheta},
\end{equation}
%
where $q:= e^{2\pi i \tau}$ and $\Theta_{m,K}(\tau,z):=\sum_{n\in {\bf Z}}
q^{K\left(n+\frac m{2K}\right)^2} e^{2\pi i z K\left(n+\frac m{2K}\right)}$.
Note that if $n,w\in {\bf Z}$, both of $m$ and $\tilde m$ must be either in ${\bf Z}$ 
or in ${\bf Z}+\frac12$ because $m\pm\tilde{m}$ must be an integer when $k=1$.
But as we see below,
in order to obtain a supersymmetric partition function, $m$ and $\tilde m$ must 
be allowed to take independent values, so that $n$ and $w$ must be allowed to 
take values in ${\bf Z}+\frac12$ as well as in ${\bf Z}$. 
In this paper we assume this to be the case. 

The first thing we notice in (\ref{thetatheta}) is that the level-1 theta functions are 
precisely the ones to construct a modular invariant partition function on the conifold
\cite{Mizoguchi,ES1,Murthy} which contains only the continuous series of $SL(2,{\bf R})$ 
(Precisely speaking, the even $k_{\mbox{\scriptsize min}}$ case 
is subtle \cite{Mizoguchi:Localized} because some lower ends of the 
continuous spectra reach the boundary of the unitary region. 
If $k_{\mbox{\scriptsize min}}$ is odd, 
the lower bound is always above the boundary. 
See FIG.1.):
\begin{equation}
Z_{\mbox{\scriptsize conifold}}^{(\mbox{\scriptsize old})}
=\int\frac{d^2\tau}{(\mbox{Im}\tau)^2}
	\frac1{(\mbox{Im}\tau)^{\frac32}\left|\eta(\tau)\right|^6}
\frac{\left|\Lambda_1(\tau)\right|^2+\left|\Lambda_2(\tau)\right|^2}{|\eta^3(\tau)|^2},
\label{Z_conifold_old}
\end{equation}
where
\begin{eqnarray}
\Lambda_1(\tau):=\Theta_{1,1}(\tau,0)
\left(\vartheta_3^2
+\vartheta_4^2
\right)(\tau,0)
-\Theta_{0,1}(\tau,0)
\vartheta_2^2(\tau,0), \\
\Lambda_2(\tau):=\Theta_{0,1}(\tau,0)
\left(\vartheta_3^2
-\vartheta_4^2
\right)(\tau,0)
-\Theta_{1,1}(\tau,0)
\vartheta_2^2(\tau,0).
\end{eqnarray}
Motivated by this observation, we define 
\begin{eqnarray}
\hat\Lambda_1(\tau,z)&:=&\Theta_{1,1}(\tau,z)
\Big(\vartheta_3(\tau,z)\vartheta_3(\tau,0)
+\vartheta_4(\tau,z)\vartheta_4(\tau,0)\Big)\nonumber\\
&&-\Theta_{0,1}(\tau,z)
\;\vartheta_2(\tau,z)\vartheta_2(\tau,0),
\label{hatLambda1}
\\
\hat\Lambda_2(\tau,z)&:=&\Theta_{0,1}(\tau,z)
\Big(\vartheta_3(\tau,z)\vartheta_3(\tau,0)
-\vartheta_4(\tau,z)\vartheta_4(\tau,0)\Big)\nonumber\\
&&-\Theta_{1,1}(\tau,z)
\;\vartheta_2(\tau,z)\vartheta_2(\tau,0)
\label{hatLambda2}
\end{eqnarray}
and 
write
\begin{eqnarray}
Z_{\mbox{\scriptsize conifold}}^{(\mbox{\scriptsize new})}(\tau)&=&
\int_0^1 ds_1 \int_0^1ds_2 \sqrt{\tau_2} (q\overline{q})^{\frac{s_1^2}4}
\frac{\left|\hat\Lambda_1(\tau,s_1\tau -s_2) \right|^2 + \left|\hat\Lambda_2(\tau,s_1\tau -s_2) \right|^2}
{|\eta(\tau)|^2 |\vartheta_1(\tau,s_1\tau-s_2)|^2}.\label{Z_conifold_new(tau)}
\end{eqnarray}
If we compare (\ref{Z_superSL2RoverU1})(\ref{thetatheta}) with 
(\ref{hatLambda1})(\ref{hatLambda2}), we see that 
$Z_{\mbox{\scriptsize conifold}}^{(\mbox{\scriptsize new})}(\tau)$ is a partition function 
of the $SL(2,{\bf R})/U(1)$ Kazama-Suzuki model coupled to a complex fermion, with a 
GSO projection performed at the stage before the Liouville limit is taken.
In fact, we can show that
$Z_{\mbox{\scriptsize conifold}}^{(\mbox{\scriptsize new})}(\tau)$ 
(i) is modular invariant, (ii) reduces to (\ref{Z_conifold_old}) if, 
after a certain regularization \cite{MOS},  
divided by a divergent volume factor and 
(iii) also contains contributions from the discrete series representations of $SL(2,R)$.

Let us first prove the modular S-invariance. In general, we can show the following equation:
\begin{eqnarray}
\Theta_{m,K}\left(
\tau,\frac{s_1\tau-s_2}{-a}
\right)
\stackrel{\tau\rightarrow -\frac1\tau}{\rightarrow}
e^{\frac{Ki\pi}{2a^2} \left(
\frac{s_1^2}\tau +\tau(1-s_2)^2 +2(s_1s_2-s_1)
\right)}
\nonumber\\
\cdot
\sqrt{\frac\tau{2Ki}}
\sum_{m'\in{\bf Z}_{2K}} e^{-\frac{mm'}k \pi i}
\Theta_{m'+\frac  K a ,K}\left(\tau,\frac{(1-s_2)\tau-s_1}{-a}
\right)
\nonumber\\
\label{S-transformed_Theta}
\end{eqnarray}
for any divisor $a$.
In comparison with the case without $\tau$-dependences through $z$,
(\ref{S-transformed_Theta}) has additional changes from (1) the exponential 
factor (the 1st line) (2) the replacement $(s_1,s_2) \rightarrow (1-s_2,s_1)$
and (3) the shift of $m'$ of $\Theta$ by an amount of $\frac K a$.

Let us now modular S-transform 
$Z_{\mbox{\scriptsize conifold}}^{(\mbox{\scriptsize new})}(\tau)$ 
(\ref{Z_conifold_new(tau)}). First, we can see that the exponential factors 
from $\hat\Lambda$'s and $\vartheta_1$ are precisely canceled by the change of 
$(q\bar{q})^{\frac{s_1^2}4}$. 
Also, the replacement $(s_1,s_2) \rightarrow (1-s_2,s_1)$  acts on 
$Z_{\mbox{\scriptsize conifold}}^{(\mbox{\scriptsize new})}(\tau)$
trivially. So all we have to do is examine the effect of the index shifts 
in the theta functions. It turns out that they are also simple because they just induce a permutation 
of the two $\hat\Lambda$'s. Thus, by counting the numbers of theta's and eta's, 
we find that  $Z_{\mbox{\scriptsize conifold}}^{(\mbox{\scriptsize new})}(\tau)$ 
(\ref{Z_conifold_new(tau)}) is modular S-invariant if 
$Z_{\mbox{\scriptsize conifold}}^{(\mbox{\scriptsize old})}$ 
(\ref{Z_conifold_old}) is. The latter statement was proven in \cite{Mizoguchi}, 
which completes the proof of the modular S-invariance of 
$Z_{\mbox{\scriptsize conifold}}^{(\mbox{\scriptsize new})}(\tau)$.

Next we turn to the modular T-invariance. Since the equations
$\hat\Lambda_1(\tau+1,z)=i\hat\Lambda_1(\tau,z)$,
$\hat\Lambda_1(\tau+1,z)=-\hat\Lambda_2(\tau,z)$
hold independently of $z$, 
we have only to worry about the change of $s_1\tau-s_2$, 
which amounts to change of variables $(s_1, s_2)\rightarrow (s_1, s_2-s_1)$.
Fortunately, the integrand of $Z_{\mbox{\scriptsize conifold}}^{(\mbox{\scriptsize new})}(\tau)$ 
(\ref{Z_conifold_new(tau)}) is periodic (with a period of 1) in $s_2$, so the integral is invariant 
under the change of variables. Thus $Z_{\mbox{\scriptsize conifold}}^{(\mbox{\scriptsize new})}(\tau)$ 
is also T-invariant.

Taking into account the four-dimensional bosons,
the total modular invariant partition function of type II strings on the conifold is now  given by 
\begin{eqnarray}
Z_{\mbox{\scriptsize conifold}}^{(\mbox{\scriptsize new})}
&=&\int\frac{d\tau d\bar\tau}{\tau_2}\frac1{\tau_2^2|\eta^2(\tau)|^2} 
Z_{\mbox{\scriptsize conifold}}^{(\mbox{\scriptsize new})}(\tau).
\end{eqnarray}



We will now 
consider the ADE generalizations of the conifold,
which are described by a coupling to an ADE modular invariant 
$N=2$ minimal model. 
We show only the result. 
For type II strings the modular invariant partition function is given by
\begin{eqnarray}
Z_{\mbox{\scriptsize ADE}}^{(\mbox{\scriptsize new})}(\tau)
& =&
\int_0^1  ds_1  \int_0^1 ds_2 \sqrt{\frac{\tau_2}k} (q\overline{q})^{\frac{ks_1^2}4}
\nonumber\\
&&\cdot
\sum_{l,\tilde{l}}N_{l,\tilde{l}} \sum_{r\in{\bf Z}_{k_{\mbox{\tiny min}}+4}+\frac l2}
\frac{
\hat{F}_{l,2r}(\tau,s_1\tau -s_2)\overline{\hat{F}_{\tilde{l},2r}(\tau,s_1\tau -s_2)}
}
{|\eta(\tau)|^2 |\vartheta_1(\tau,s_1\tau-s_2)|^2},
\label{Z_ADE_new(tau)}
\end{eqnarray}
\begin{widetext}
\begin{eqnarray}
\hat{F}_{l,2r}(\tau,z)&=&\frac14 
\sum_{m\in {\bf Z}_{4(k_{\mbox{\tiny min}}+2)}} 
\left[
\vartheta_3(\tau,0)\vartheta_3(\tau,z)\mbox{ch}^{\mbox{\scriptsize NS}}_{l,m}(\tau,0)
-(-1)^{r-\frac m2}
\vartheta_4(\tau,0)\vartheta_4(\tau,z)\mbox{ch}^{\widetilde{\mbox{\scriptsize NS}}}_{l,m}(\tau,0)
\right.
\nonumber\\
&&\left.
-\vartheta_2(\tau,0)\vartheta_2(\tau,z)\mbox{ch}^{\mbox{\scriptsize R}}_{l,m}(\tau,0)
\right]
\cdot
\Theta_{
(k_{\mbox{\scriptsize min}}+2)2r-(k_{\mbox{\scriptsize min}}+4)m,
2(k_{\mbox{\scriptsize min}}+2)(k_{\mbox{\scriptsize min}}+4)
}
\left({\textstyle
\tau,\frac z{k_{\mbox{\scriptsize min}}+4}
}\right).
\nonumber\\
\label{hatF}
\end{eqnarray}
\end{widetext}
$N_{l,\tilde l}$ is the coefficients of the ADE modular invariant. 
We have used the same notation for $N=2$ minimal characters as
in \cite{ES1}.
We note the particular $z$-dependence of $\hat{F}_{l,2r}(\tau,z)$ (\ref{hatF}), 
which is similar to $F_{l,2r}(\tau)$ defined in \cite{ES1}. 
The proof of the modular invariance is parallel to that in 
the conifold case. The only nontrivial point is the $\tau$-dependence 
through the $z$-argument. In the present case the modular S-transformation 
simply permutes $\hat F$'s cyclically, and  
$Z_{\mbox{\scriptsize ADE}}^{(\mbox{\scriptsize new})}(\tau)$ 
as a whole remains invariant. The proof of the modular T-invariance is also 
similar.

Conversion to heterotic string theories  is also straightforward.
Since the transverse fermion theta's have no $z$-dependence, all we 
need to do is replace \cite{Gepner} 
$\frac{\vartheta_3\pm\vartheta_4}{\eta}$
by
$\frac{\vartheta_3^5\mp\vartheta_4^5}{\eta^5}\cdot
\frac{\vartheta_3^8+\vartheta_4^8}{\eta^8}$, and
$\frac{\vartheta_2\pm\tilde\vartheta_1}{\eta}$
by
$-\frac{\vartheta_2^5\pm\tilde\vartheta_1^5}{\eta^5}\cdot
\frac{\vartheta_2^8+\tilde\vartheta_1^8}{\eta^8}$ for $E_8\times E_8$
($\tilde\vartheta_1(\tau,z):= (\Theta_{1,2}-\Theta_{-1,2})(\tau,z)$)
(for the transverse fermion theta's only), and in an analogous fashion for 
$SO(32)$.
Then the partition functions 
remain modular invariant.

\section{Localized modes}

We will now extract the discrete spectrum by using the technique developed in 
\cite{MOS,HPT,ES2}.
We first express various theta functions in 
$Z_{\mbox{\scriptsize ADE}}^{(\mbox{\scriptsize new})}(\tau)$
in terms of traces of some operators over appropriate Hilbert spaces. 
We define ($y:= e^{2\pi i z}$):
\begin{itemize}
\item{
${\cal H}^{SL(2,{\bf R})}_{\pm,(h,l_0)} $ as an $SL(2,{\bf R})$ current algebra module 
generated from a state $|h,l_0\rangle$ such that 
\begin{eqnarray}
&&L^{SL(2,{\bf R})}_0|h,l_0\rangle= h|h,l_0\rangle,~~~
J_0^3|h,l_0\rangle= l_0|h,l_0\rangle,\nonumber\\
&&L^{SL(2,{\bf R})}_n|h,l_0\rangle=J_n^3|h,l_0\rangle=J_n^+|h,l_0\rangle=J_n^-|h,l_0
\rangle=0
\nonumber\\&&
~~~(n>0)~~~
\mbox{and}~~~J_0^\mp|h,l_0\rangle=0,\nonumber
\end{eqnarray}
satisfying
$\mbox{Tr}_{{\cal H}^{SL(2,{\bf R})}_{\pm,(h,l_0)}}q^{L_0^{SL(2,{\bf R})}}y^{J_0^3}=
\frac{\pm i q^{\frac18 + h} y^{\mp\frac12 + l_0}}{\vartheta_1(\tau,z)}$.
}
\item{${\cal H}_{\nu,2}$ ($\nu\in{\bf Z}_4$)
as a free fermion module such that 
$\mbox{Tr}_{{\cal H}_{\nu,2}}q^{L_0^{(\nu)}}y^{F^{(\nu)}}=
q^{\frac1{24}}\frac{\Theta_{\nu,2}(\tau,z)}{\eta(\tau)}$,
where if $\nu=0,2$, $L_0^{(\nu)}$ and $F^{(\nu)}$ denote
the Virasoro $L_0$ and fermion number operators in the NS sector, 
while  if $\nu=\pm1$, they are the ones in the R sector.
}
\item{${\cal H}_{m,K}$ ($m\in{\bf Z}_{2K}$)
as a free boson module such that 
$\mbox{Tr}_{{\cal H}_{m,K}}q^{L_0^{U(1)}}y^{J_0^{U(1)}}=
q^{\frac1{24}}\frac{\Theta_{m,K}(\tau,z)}{\eta(\tau)}$,
where $J_0^{U(1)}$ is the $U(1)$ charge operator.}
\item{${\cal H}^{l,s}_{m}$ 
as an irreducible module of \mbox{$N=2$} minimal superconformal algebra
such that
$\mbox{Tr}_{{\cal H}^{l,s}_{m}}q^{L_0^{N=2}}y^{J_0^{N=2}} = 
\chi^{l,s}_m(\tau,z)
$. }
\end{itemize}
For convenience we also define
\begin{eqnarray}
{\cal H}^{(\nu)}_{F_{l,2r}}:=
							{\oplus}_{\mbox{\tiny $
							\begin{array}{c}
							\nu_0,\nu_1,\nu_2\in Z_2\\
							\nu_0+\nu_1+\nu_2 \\
							\equiv r-\frac l2({\rm mod}2)
							\end{array}$}
							}
											\left(
					{\cal H}_{\nu+2\nu_0,2}
					\otimes {\cal H}_{\nu+2\nu_1,2}
					\otimes {\cal H}_{\nu}^{l,\nu+2\nu_2}
				\right)\otimes 
				{\cal H}
				_{(k_{\mbox{\scriptsize min}}+2)2r
				-(k_{\mbox{\scriptsize min}}+4)(l+\nu),
									2(k_{\mbox{\scriptsize min}}+2)
									(k_{\mbox{\scriptsize min}}+4)
				}.\nonumber\\
												\end{eqnarray}
We can then write 
(for diagonal 
modular invariants for simplicity)
\begin{widetext}
\begin{eqnarray}
&&\int_0^1 ds_1 \int_0^1 ds_2 
\sqrt{\frac{\tau_2}k} (q\bar q)^{\frac{ks_1^2}4}
\left|
\frac{\hat F_{l,2r}(\tau, s_1\tau - s_2)}{\eta^3(\tau) \vartheta_1(\tau, s_1\tau - s_2)}
\right|^2
\nonumber\\
&=&
\int_0^1 ds_1 \int_0^1 ds_2 
\sqrt{\frac{\tau_2}k} (q\bar q)^{\frac{ks_1^2}4}
\frac14 \sum_{\nu,\tilde{\nu}\in{\bf Z}_{4(k_{\mbox{\tiny min}}+2)}}(-1)^{\nu+\tilde{\nu}}
\mbox{Tr}_{\left( {\cal H}^{SL(2,{\bf R})}_{+,(0,0)} 
	\otimes{\cal H}_{F_{l,2r}}^{(\nu)}
	\right)
	\otimes
	 \left({\cal H}^{SL(2,{\bf R})}_{+,(0,0)} 
	\otimes{\cal H}_{F_{l,2r}}^{(\tilde\nu)}
	\right)}
\nonumber\\
&&\cdot q^{L_0^{SL(2,{\bf R})}+L_0^{(\nu)}+L_0^{(\nu)}+L_0^{U(1)}
	\!\!\!-\frac14-\frac{c_{\mbox{\tiny min}}}{24}
	+s_1(J_0^3+F^{(\nu)}+\frac{J_0^{U(1)}}{k_{\mbox{\tiny min}}+4}+\frac12)}	
	\nonumber\\
&&
\cdot 
\bar{q}^{\tilde{L}_0^{SL(2,{\bf R})}+\tilde{L}_0^{(\tilde{\nu})}+\tilde{L}_0^{(\tilde{\nu})}+
 	\tilde{L}_0^{U(1)}
	\!\!\!-\frac14-\frac{c_{\mbox{\tiny min}}}{24}
	+s_1(\tilde{J}_0^3+\tilde{F}^{(\tilde{\nu})}+\frac{\tilde{J}_0^{U(1)}}{k_{\mbox{\tiny min}}+4}
	+\frac12)}	\nonumber\\
&&\cdot e^{-2\pi i s_2 (J_0^3+F^{(\nu)}+\frac{J_0^{U(1)}}{k_{\mbox{\tiny min}}+4}
-\tilde{J}_0^3-\tilde{F}^{(\tilde{\nu})}-\frac{\tilde{J}_0^{U(1)}}{k_{\mbox{\tiny min}}+4})}. \label{to_do_s1s2integrals}
\end{eqnarray}
%
The $s_2$ integral yields a constraint $J^3_0+F^{(\nu)}
+\frac{J_0^{U(1)}}{k_{\mbox{\scriptsize min}}+4}
=\tilde{J}^3_0+\tilde{F}^{(\tilde{\nu})}
+\frac{\tilde{J}_0^{U(1)}}{k_{\mbox{\scriptsize min}}+4}
(:= J_0^{\mbox{\scriptsize tot}})$.  
Using the Fourier transformation
$\sqrt{\frac{\tau_2}k}(q\bar q)^{\frac{ks_1^2}4}
=\frac1k\int_{-\infty}^\infty dc ~e^{-\frac \pi{k\tau_2} c^2 -2 \pi i c s_1}
$,
the $s_1$ integral can also be done. 
We further use isomorphisms of various Hilbert spaces 
 (spectral flows with respect to 
$J_0^{\mbox{\scriptsize tot}}$) to find
($c=2\tau_2 p$, $\kappa=k+2$)
\begin{eqnarray}
(\ref{to_do_s1s2integrals})
&=&\frac1{4k} \sum_{\nu,\tilde{\nu}\in{\bf Z}_{
4(k_{\mbox{\tiny min}}+2)
}}(-1)^{\nu+\tilde{\nu}}
 \left[
 \mbox{Tr}_{\left( {\cal H}^{SL(2,{\bf R})}_{-,(-\frac\kappa 4,-\frac\kappa 2)} 
	\otimes 
	{\cal H}^{(\nu)}_{F_{l,2(r+1)}}
	\right)
	\otimes
	 \left({\cal H}^{SL(2,{\bf R})}_{-,(-\frac\kappa 4,-\frac\kappa 2)} 
	 {\cal H}^{(\tilde\nu)}_{F_{l,2(r+1)}}
	\right)}
	\right.
\nonumber\\
&&
\cdot \int_{-\infty}^\infty dp 
\frac{q^{\frac1k(p+\frac{ik}2)^2-\frac14 -\frac{c_{\mbox{\tiny min}}}{24}
+L_0^{SL(2,{\bf R})}+2L_0^{(\nu)}
+L_0^{N=2}+L_0^{U(1)}+\frac\kappa 4
}
\bar{q}^{\frac1k(p+\frac{ik}2)^2-\frac14 -\frac{c_{\mbox{\tiny min}}}{24}
+\tilde{L}_0^{SL(2,{\bf R})}+2\tilde{L}_0^{(\tilde{\nu})}
+\tilde{L}_0^{N=2}
+\tilde{L}_0^{U(1)}+\frac\kappa 4
}}
{-2\pi(ip+J_0^{\mbox{\scriptsize tot}}+\frac12)}
\nonumber\\
&&
-\mbox{Tr}_{\left( {\cal H}^{SL(2,{\bf R})}_{+,(0,0)} 
	\otimes 
	{\cal H}^{(\nu)}_{F_{l,2r}}
	\right)
	\otimes
	 \left({\cal H}^{SL(2,{\bf R})}_{+,(0,0)} 
	\otimes 
	{\cal H}^{(\tilde\nu)}_{F_{l,2r}}
	\right)}
\nonumber\\
&&
\left.
\cdot 
\int_{-\infty}^\infty dp 
\frac{q^{\frac{p^2}k-\frac14 -\frac{c_{\mbox{\tiny min}}}{24}
+L_0^{SL(2,{\bf R})}+2L_0^{(\nu)}+L_0^{N=2}+L_0^{U(1)}
}
\bar{q}^{\frac{p^2}k-\frac14 -\frac{c_{\mbox{\tiny min}}}{24}
+\tilde{L}_0^{SL(2,{\bf R})}+2\tilde{L}_0^{(\tilde{\nu})}+\tilde{L}_0^{N=2}+\tilde{L}_0^{U(1)}
}}
{-2\pi(ip+J_0^{\mbox{\scriptsize tot}}+\frac12)}
\right].
\end{eqnarray}
\end{widetext}

Thanks to the isomorphisms 
we have used, the Hilbert space of the first trace has been changed from 
that of $F_{l,2r}$ to $F_{l,2(r+1)}$. 
As was done in \cite{MOS,HPT,ES2}, we will now change the integration contour 
of the first trace from $p':= p+\frac {ik}2 \in {\bf R}+\frac {ik}2$ to ${\bf R}$.
Then it picks up a residue of the pole at $p=i(J_0^{\mbox{\scriptsize tot}}+\frac12)$  for the states satisfying 
$-\frac{k+1}2<J_0^{\mbox{\scriptsize tot}}<-\frac12$. These negative-momentum 
states reside below the lower bound of the continuous spectrum, and precisely on 
the boundary of the unitary region \cite{BFK,DLP}. 

\begin{figure}
\includegraphics[width=150mm]{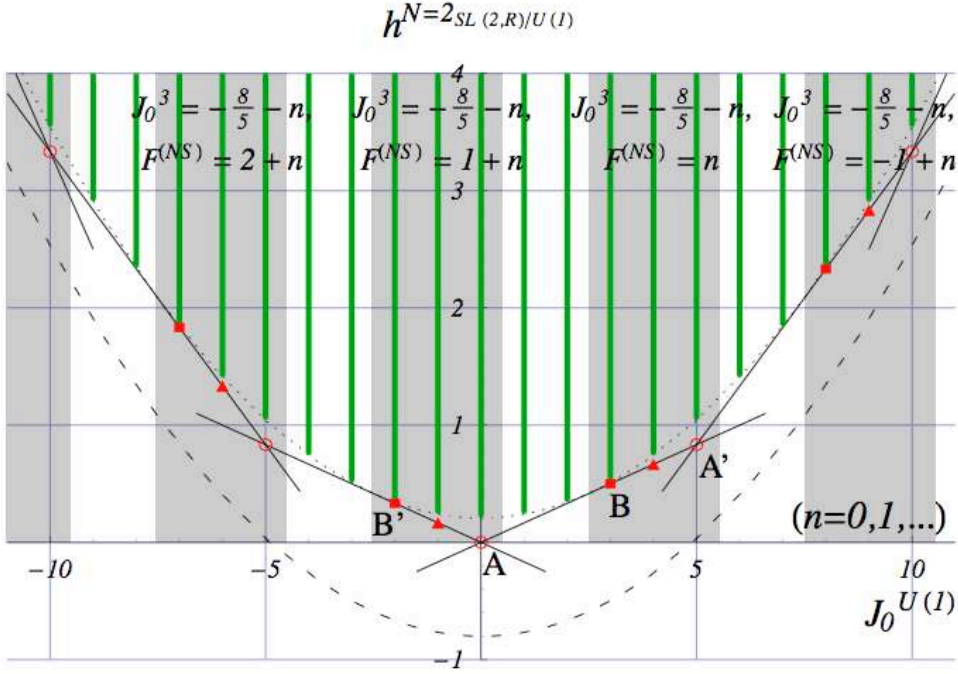}
\caption{\label{fig} The unitary region of the $c_{KM}=8$ ($k_{\mbox{\scriptsize min}}=1$),
$N=2$ superconformal algebra (NS sector). 
Green and red indicate, respectively, the continuous and 
discrete series representations contained in the spectrum. 
}
\end{figure}

We illustrate this in an example. 
FIG.1 shows the discrete-series representations of $N=2$ superconformal algebra
associated with the noncompact coset for $k=\frac65$ ($k_{\mbox{\scriptsize min}}=1$) 
in the NS-sector. The green lines are the continuous spectrum contained in the partition 
function, and the red points on the segments are the discrete states coming from the 
pole contributions. For the type II case, the points A and B  only give massless states.
They come from poles in the orbits $F_{0,0}$ and $F_{0,2}$ (after the isomorphisms),
and with their superpartners they constitute a single vector- of a hyper-multiplet depending 
on the chirality. For the ($E_8\times E_8$) heterotic case, there are two additional points,
A' and B', in the left mover which contribute to massless states. Various conformal weights 
of states which couple to these discrete states are shown in TABLE.I. It turns out 
that A (if paired with A in the right mover) gives $10\oplus 1$ and A' does a singlet of 
$SO(10)$. With 16 states coming from the left Ramond sector, they constitute  
real scalars in the $27\oplus 1$ of $E_6$. B ,B' and their corresponding left Ramond states 
yield another $27\oplus 1$. Ttaking into account their superpartners and the 
symmetry $\hat F_{l,2r}=\hat F_{k_{\rm min}-l,2r+k_{\rm min}+4}$,
we find two ${\cal N}=1$ massless scalar multiplets in the $27\oplus 1$ of 
$E_6$ localized at the cigar tip.

No massless graviton 
is localized (nor massless gauge fields in the heterotic case), but
if the noncompact Calabi-Yau is not the whole internal manifold itself 
but a part of some finite-size but large manifold compared to the scale of 
the collapsing cycles, which are isolated and distant from the rest,
then there should be a massless graviton (and gauge fields for heterotic strings) 
associated with the constant mode, 
and they will interact with the localized fields. We hope that the 
singular CFT compactifications we presented here will lead to a new, 
interesting realization of local $E_6$ GUT with less moduli in this simple setting.  

\begin{table}
\caption{\label{tab:table1}Breakdowns of conformal weights 
constituting massless states for the $E_8\times E_8$ heterotic
string
($k_{\mbox{\scriptsize min}}=1
$, the left NS-sector).}
\begin{ruledtabular}
\begin{tabular}{ccccc}
Rep.&A                   & A'     & B         & B'  \\
\hline
Lower bound&$0+\frac5{24}$
		    &$\frac56+\frac5{24}$
		    &$\frac3{10}+\frac5{24}$
		    & $\frac2{15}+\frac5{24}$  \\
$N=2$ minimal& $0$
  			   &$\frac16$
			   &$0$
			   &$\frac16$ \\
Imaginary momentum factor &$-\frac5{24}$
  						  &$-\frac5{24}$
						  &$-\frac1{120}$
						  &$-\frac1{120}$ \\
  
$SO(10)$ fermions &$\frac12$ or $0$
  		 		  &$0$
				  &$\frac12$ or $0$
				  &$0$\\
Liouville fermions & $\frac12$ or $0$       
  				& $0$  
				& $0$ or $\frac12$
				&$\frac12$ \\
$L_0^{SL(2,{\bf R})}$ & $0$  or $1$   
  				     & $0$  
				     &$0$ 
				     &$0$ \\
 \hline
  Total & $1$&$1$&$1$&$1$\\
   \hline
  $SO(10)$ representation &${\bf 10}\oplus{\bf 1}$
  					   &${\bf 1}$
					   &${\bf 10}\oplus{\bf 1}$
					   &${\bf 1}$\\
 \end{tabular}
\end{ruledtabular}
\end{table}

\begin{acknowledgments}
The author wishes to thank T.~Eguchi and Y.~Sugawara for valuable discussions. 
The author also thanks M.~Hatsuda, T,~Kawai, H.~Kodama, N.~Ohta, Y.~Okada, 
Y.~Satoh and I.~Tsutsui for discussions. %
This work is supported in part by the Grand-in-Aid
for Scientific Research No.18540287.


\end{acknowledgments}

\end{document}